\documentclass[twocolumn,aps,prl,superscriptaddress,tightenlines]{revtex4-1}

\usepackage{amsmath, amsfonts, amsthm, amssymb, graphicx, xcolor}
\usepackage{hyperref}
\hypersetup{
    colorlinks,
    linkcolor={red!75!black},
    citecolor={blue!75!black},
    urlcolor={blue!75!black}
}
\usepackage{subfigure}

\allowdisplaybreaks[3]
\def \bal#1\eal  {\begin{align} #1 \end{align}}
\newcommand{\be} {\begin{equation}}
\newcommand{\ee} {\end{equation}}
\newcommand{\nn} {\nonumber}

\newcommand{\mn} {{\mu\nu}}

\def\ba{\begin{eqnarray}}
\def\ea{\end{eqnarray}}

\def\L{\mathcal{L}}

\def\mn{_{\mu \nu}}

\def\({\left(}
\def\){\right)}

\def\mpl{M_{\rm Pl}}
\def\p{\partial}
\def\ie{{\em i.e. }}

\begin{document}

\title{Gravitational Rainbows: LIGO and Dark Energy at its Cutoff}
\author{Claudia de Rham}
\email{c.de-rham@imperial.ac.uk}
\affiliation{Theoretical Physics, Blackett Laboratory, Imperial College, London, SW7 2AZ, U.K. }
\date{\today}
\affiliation{CERCA, Department of Physics, Case Western Reserve University, 10900 Euclid Ave, Cleveland, OH 44106, USA}
\author{Scott Melville}
\email{s.melville16@imperial.ac.uk}
\affiliation{Theoretical Physics, Blackett Laboratory, Imperial College, London, SW7 2AZ, U.K. }
\date{\today}

\begin{abstract}
\noindent The recent direct detection of gravitational waves from a neutron star merger with optical counterpart has been used to severely constrain models of dark energy that typically predict a modification of the gravitational wave speed. However, the energy scales observed at LIGO, and the particular frequency of the neutron star event, lie very close to the strong coupling scale or cutoff associated with many dark energy models. While it is true that at very low energies one expects gravitational waves to travel at a speed different than light in these models, the same is no longer necessarily true as one reaches energy scales close to the cutoff. We show explicitly how this occurs in a simple model with a known partial UV completion. Within the context of Horndeski, we show how the operators that naturally lie at the cutoff scale can affect the speed of propagation of gravitational waves and bring it back to unity at LIGO scales. We discuss how further missions including LISA and PTAs could play an essential role in testing such models.
\end{abstract}

\maketitle

%%%%%%%%%%%%%%%%
%Introduction
%%%%%%%%%%%%%%%%

\noindent{\bf Dark Energy after GW170817 and GRB170817A:}
The recent direct detections of gravitational waves (GWs) have had an unprecedented impact on our understanding of gravity at a fundamental level.
The first event alone (GW150914 \cite{Abbott:2016blz}) was already sufficient to put bounds on the graviton with better precision than what we know of the photon.
Last year, the first detection of GWs from a neutron star merger (GW170817), some $10^{15}$ light seconds away, which arrived within one second of an optical counterpart (GRB170817A), allowed us to constrain the GW speed with remarkable precision \cite{TheLIGOScientific:2017qsa, Monitor:2017mdv, GBM:2017lvd}
\ba
\label{eq:aLIGOconstraint}
-3\times 10^{-15}\le \frac{c_T}{c_\gamma}-1\le 7 \times 10^{-16}\,,
\ea
with $c_T$ the GW phase velocity and $c_\gamma$ the speed of light.  \\

Such a constraint has had far-reaching consequences for models of dark energy.
Within the context of the Effective Field Theory (EFT) for dark energy \cite{Gubitosi:2012hu}, it was rapidly pointed out that \eqref{eq:aLIGOconstraint} was sufficient to suppress the EFT operators that predict non-luminal gravitational propagation\cite{Lombriser:2015sxa,Lombriser:2016yzn,Creminelli:2017sry, Sakstein:2017xjx, Ezquiaga:2017ekz, Baker:2017hug, Akrami:2018yjz, Heisenberg:2017qka, BeltranJimenez:2018ymu}. In particular, within the framework  of scalar-tensor  theories of gravity, Horndeski \cite{Horndeski:1974wa} has played a major part in the past decade as a consistent ghost-free EFT in which the scalar degree of freedom could play the role of dark energy. Yet the interplay between the scalar and gravity typically implies that GWs would not travel luminally. The LIGO constraint on the GW speed only leaves out the generalization of the cubic Galileon \cite{Deffayet:2010qz},
which is severely constrained by other observations. As a result the Horndeski EFT seems almost entirely ruled out as a dark energy candidate \footnote{
The impact of \eqref{eq:aLIGOconstraint} is not limited to scalar-tensor theories of gravity---other models of dark energy, such as vector-tensor gravity or scalar-vector-tensor gravity, have also seen their parameter space remarkably affected (see however \cite{Heisenberg:2017qka,BeltranJimenez:2018ymu,Belgacem:2017cqo} for models that survive the bound). Lorentz-violating theories have also been profoundly constrained \cite{Mirshekari:2011yq,Yunes:2016jcc,Blas:2016qmn}, although we shall focus on theories which are fundamentally LI here}.  \\

%%%%%%%%%%%%%%%%%%%%%%%%%%%%%%%%
\begin{figure}[h!]
\includegraphics[width=0.5\textwidth]{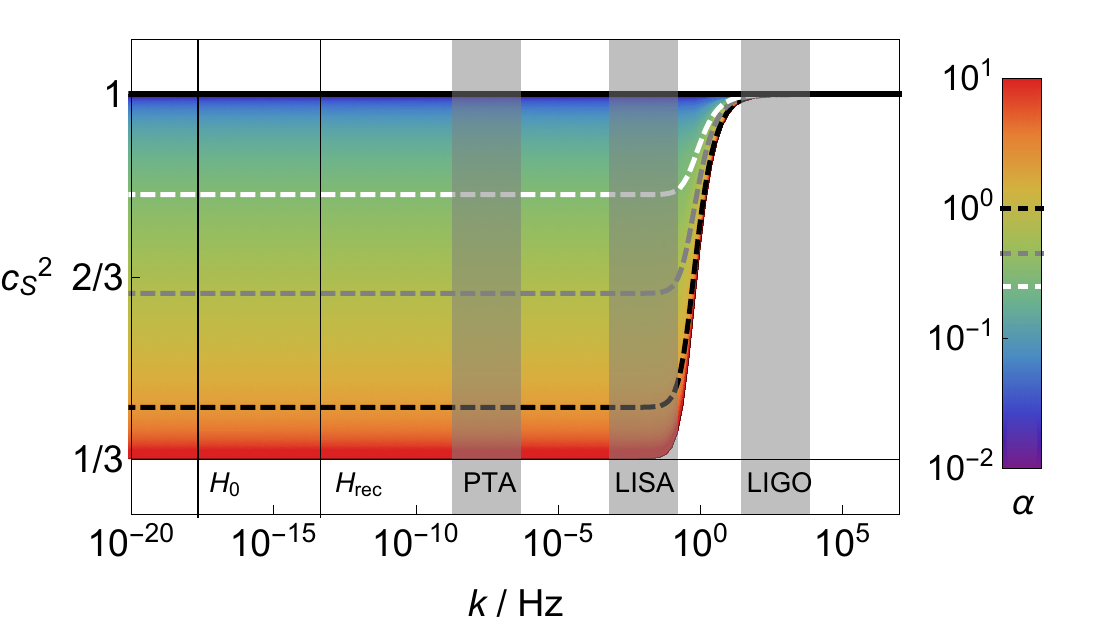}
\caption{Sound speed for $\delta \phi$ fluctuations in the model \eqref{eqn:UVscalar}. While subluminal at  $k\ll M$, luminality is recovered above the cutoff (here, $M=10^{-3} \Lambda$). The EFT can safely describe cosmology from today $H_0$ to before recombination $H_{\rm rec}$, but may receive order one corrections in the LIGO band.
%The frequency sensitivities of Pulsar Timing Arrays ($\sim 10^{-9}$--$10^{-7}$Hz), LISA ($\sim 10^{-4}$--$10^0$Hz) and LIGO ($\sim 10^1$--$10^{3}$Hz).
%In this realization, the cutoff of the partial UV completion is as high as $100 \Lambda \sim 10^4$Hz, and can be trusted at LIGO frequencies.
\label{fig:example}}
\end{figure}
%\vspace{1cm}
%%%%%%%%%%%%%%%%%%%%%%%%%%%%%%%%

%In what follows we shall not discuss any further which particular models may or may not be consistent with the recent bound \eqref{eq:aLIGOconstraint}, but rather emphasize the point that
Nevertheless, it should be noted that the recent LIGO bound applies to GWs at a frequency of $10-100$Hz, while the EFT for dark energy is ``constructed'' as an effective field theory for describing cosmology on scales 20 orders of magnitude smaller.
When it comes to constraining such EFT parameters, it is therefore important to recall that they could in principle depend on scale: generically, the GW speed may depend on the frequency at which it is measured,
$c_T=c_T(k)$. The LIGO bound \eqref{eq:aLIGOconstraint} should therefore be read as a constraint on $c_T(k)$ at frequencies on the order of $k\sim 10-100$Hz, and from their very construction we expect EFTs such as Horndeski to break down at a cutoff $\sim 100$Hz if not much lower. If the theory is to ever admit a Lorentz-invariant (LI) high energy (UV) completion, then the front velocity \footnote{Strictly speaking, the front velocity is defined as the speed of the front of a disturbance \cite{Brillouin,Milonni}. In practice, the front velocity is defined as the high frequency limit of the phase velocity \cite{Brillouin,Milonni,deRham:2014zqa},
\ba
c_{\rm front}=\lim_{k\to \infty}c_T(k)\,.
\ea}
{\bf must be luminal} which implies that the sound speed $c_T(k)$ {\bf will necessarily asymptote to exactly luminal} at high frequencies. While the EFT of dark energy may predict a GW sound speed that departs from unity at low energy, it is nonetheless natural to expect a speed arbitrarily close to luminal at higher frequencies.
%Precisely what we mean by ``higher frequencies" depends on the details of the UV completion.
In the case of Horndeski, the scale of the cosmological background generally requires that new physics ought to enter at (or  parametrically before) the energy scales observed at LIGO, where it would be natural to observe a luminal velocity.
We shall present how this would naturally occur in a simple scalar field model (Fig.~\ref{fig:example}) before turning to the full-fledged scalar-tensor theory and discussing the implications of LI--UV completions to Horndeski. \\

%%%%%%%%%%%%%%%%
\noindent{\bf Scalar EFT:}
%%%%%%%%%%%%%%%%
%
We start by looking at a simple yet representative scalar EFT example \footnote{The argument goes through essentially unaffected if instead of \eqref{eq:egPX} we had  chosen an arbitrary function $\Lambda^4 P( (\p\phi)^2/\Lambda^4)$. We chose the particular form of \eqref{eq:egPX} for concreteness.}
\ba
\mathcal{L}_{M} = -\frac 12 (\p \phi)^2 + \frac{1}{2\Lambda^4}(\p \phi)^4 + \mathcal{O} \left( \frac{ (\partial^2 \phi )^2 }{M^2}  \right) \,,
\label{eq:egPX}
\ea
where $M$ is the scale of new physics.
%In a standard EFT picture (that would not invoke a Vainshtein mechanism) we expect new physics to enter at $M \leq \Lambda$ to ``save'' the EFT from perturbative unitarity breaking that would otherwise occur at the scale $\Lambda$.
$\phi$ could be a placeholder for dark energy---for instance, let us set $\langle \phi \rangle =\alpha \Lambda^2 t$ as the background and consider fluctuations $\phi=\langle \phi \rangle +\delta \phi$.
%, where for now we do not put any a priori restriction on the dimensionless coefficient $\alpha$.
As is well known, on this  spontaneously Lorentz-breaking (sLB) background the sound speed for $\delta \phi$ is
\begin{equation}
c_S^2=1-\Delta_0=1-\frac{4 \alpha^2}{1+6 \alpha^2} \,,
% \Delta c_S^2  =  1- c_S^2 = \frac{4 \alpha^2}{1+6 \alpha^2}  = \Delta_0^2\,.
\label{eq:cS}
\end{equation}
leading to an order one deviation from luminality  if the parameter $\alpha \sim \mathcal{O} (1)$.
At this stage, we may wonder if we can trust a background configuration close to the strong coupling scale $\Lambda$. This question has been the subject of extensive work and we refer the reader to \cite{deRham:2014wfa} for careful considerations. Here, we take the approach that the EFT can be re-organized as a derivative expansion in which, while the field gradient may be ``large'', higher derivatives of the field are suppressed.
%\footnote{Strictly speaking most models of Horndeski-dark energy violate even that weaker EFT re-organization criterion, but in this paper we shall give those models the benefit of the doubts as appropriate models for dark energy and put aside the question of naturalness}.
%This re-organization is possible in this context thanks to the existence of a shift symmetry.
This means that a profile with $\dot \phi\sim \Lambda^2$ may be considered without going beyond the regime of validity of the EFT so long as higher derivatives are suppressed: $\p^n \phi \ll M^{n+1} \lesssim \Lambda^{n+1}$ for any $n\ge 2$.
Concretely, this implies that background configurations with $\alpha \sim \mathcal{O}(1)$
% (or even $|\alpha|\gg 1$)
do not necessarily lead to order one contributions from other irrelevant operators.
We follow this approach here as it is the one used in the context of Horndeski models of dark energy. \\

%%%%%%%%%%%%%%%%
\noindent{\bf EFT Cutoff:}
%%%%%%%%%%%%%%%%
%
%Having established the (potential) consistency of the background, we now turn to that of perturbations.
The model \eqref{eq:egPX} predicts a speed of sound \eqref{eq:cS} which appears to be the same irrespective of the frequency of the $\delta \phi$ fluctuations. Yet if we consider $\delta \phi$-waves at sufficiently high frequencies, they should be insensitive to the sLB background. LI should be restored \footnote{Even if the UV completion was not LI, it would be surprising that high energy physics knows about the scale of the sLB background.}
and hence high-frequency $\delta \phi$ waves should be exactly luminal. The reason this is not manifest in \eqref{eq:cS} is because we are working within the EFT \eqref{eq:egPX}, which is only consistent at frequencies much smaller than the cutoff, $M$. Interestingly, in the context of the GW170817 detection, the frequency of the GWs span from $24$Hz to a few hundred Hz,
%\ie within the range of $\sim (1-10)\times 10^{-14}$eV
which is perilously close to the strong coupling scale associated with many Horndeski dark energy models \footnote{
In particular, this is the largest strong coupling scale which would allow the second derivative operators,
$\partial^2 \phi$, to have an order unity effect on cosmological perturbations. For instance,
$ (\partial \phi )^2 \frac{\Box \phi}{\Lambda^3} \sim  (\partial \delta \phi )^2 \frac{M_P H_0^2}{\Lambda^3} $
remembering that $\dot \phi^2 \sim M_P^2 H_0^2$ if an approximately shift symmetric ($\phi \propto t $) scalar is to drive the late time expansion of the universe.
},
\ba
\label{eq:LambdaHorn}
M \lesssim \Lambda_{\rm Horndeski} \sim (\mpl H_{0}^2)^{1/3} \sim 260{\rm Hz} %\\
%&\sim&10^{-13}{\rm eV}\sim 260{\rm Hz}\,,\nn
\ea
where $H_0$ is the Hubble parameter today.
At those scales, the EFT \eqref{eq:egPX} can no longer be the appropriate description for the $\delta \phi$-waves, as we have neglected operators of the form $(\p^2 \phi)^2/M^2$, where $M$ is the cutoff \footnote{
Note that the existence of higher derivative operators in this EFT should not be confused with the existence of an Ostrogradsky ghost. Indeed, higher derivative operators naturally enter from  integrating out heavy degrees of freedom, and just manifest the fact that the EFT breaks down at the cutoff scale.
}. The existence of such higher derivative operators cannot be ignored---they are  mandated by positivity bounds if this theory is to admit a sensible Wilsonian UV completion \cite{Adams:2006sv, deRham:2017imi}.  \\

%%%%%%%%
\noindent{\bf Sound Speed near the Cutoff:}
%%%%%%%%
The low-energy EFT \eqref{eq:egPX} is appropriate when considering $\delta \phi$-waves at frequencies $k/M \ll 1$, however at higher frequencies one should include the irrelevant operators that naturally enter the  EFT at the scale $M$ and modify the dispersion relation,
\begin{equation}
c_S^2 (k) = 1- \Delta_0  +  \Delta_2   \frac{k^2}{M^2} + \mathcal{O} \left(  \frac{k^4}{M^4} \right) \,,
\label{eqn:DcsNLO}
\end{equation}
where the running $\Delta_2$ is controlled by the higher order operators.
This scale-dependence of the sound speed is unavoidable: not only are the next-to-leading order operators required in order to properly renormalize divergences within the EFT, they also naturally arise from a generic UV completion.
Of course when reaching  the scale $M$, we lose control of the EFT and the precise details of the UV completion are essential in determining the sound speed of $\delta \phi$-waves (even if---as we have argued---the background configuration itself may not be sensitive to the UV completion).  \\

To give a precise example of how  UV physics \footnote{
 $P(X)$ theories such as \eqref{eq:egPX} can be seen as the low energy EFT of a high energy $U(1)$ theory, broken to nothing as one integrates over the (massive) radial component. In that case the completion could be renormalizable and could then be a complete (rather than partial) UV completion of \eqref{eq:egPX}. For the case of Horndeski, it will certainly not be our aim to find a renormalizable completion and for simplicity we shall consider  \eqref{eq:egPX} as a potential completion here. We thank Paolo Creminelli and Filippo Vernizzi for pointing this out.
} may affect the sound speed at frequencies close to $M$, consider the following specific situation where the massless scalar $\phi$ couples to a heavy scalar $\chi$ via,
\ba
\mathcal{L}_{ \Lambda_* }  &=& -\tfrac{1}{2} (\p \phi)^2 -\tfrac12 (\p \chi)^2 -\tfrac 12 M^2 \chi^2+\frac{\chi}{\Lambda_*} (\p \phi)^2\,,
\label{eqn:UVscalar}
\ea
where $\chi$ becomes dynamical around $M$ and strongly coupled at a scale $\Lambda_* $.
For \eqref{eqn:UVscalar} to represent a (partial) completion of \eqref{eq:egPX} with an extended region of validity, we require the scale hierarchy $\Lambda_*\gg M$ implying
\ba
\label{eq:HierarchyScales}
M \ll \Lambda=(M \Lambda_*)^{1/2}\ll \Lambda_*\,.
\ea
Even though \eqref{eq:egPX} only becomes strongly coupled at the scale $\Lambda$, its cutoff is in fact even smaller $M\ll \Lambda$ (see \cite{Bellazzini:2016xrt}, this hierarchy also appears in the case of Galileons \cite{deRham:2017imi} and massive gravity \cite{Bellazzini:2017fep,deRham:2017xox}, \footnote{Note that \cite{Bellazzini:2017fep} included some confusions in interpreting the context in which EFTs should be valid that were clarified in \cite{deRham:2017xox}.}).
Integrating out $\chi$ at tree level gives the EFT \eqref{eq:egPX} with additional irrelevant operators
\ba
    \mathcal{L} = - \frac{1}{2} ( \partial \phi )^2 +  \frac{1}{2 \Lambda^4} (\partial \phi )^2 \frac{M^2}{ M^2 - \Box} ( \partial \phi )^2 \,.   \quad
\ea
Including these irrelevant operators, we find a dispersion relation
\ba
\omega^2 = k^2 - \frac{4 \alpha^2}{ 1 + 2 \alpha^2 }  \frac{ \omega^2 M^2 }{ M^2 -\omega^2 + k^2 }\,,
\ea
which matches the leading order EFT sound speed \eqref{eq:cS} at sufficiently small frequencies  $k\ll M$ but leads to luminality at higher frequencies $\omega^2=k^2\(1 + \cdots\)$, where the ellipses vanish at high energy and their precise form depends on the details of the completion. The exact behavior of the sound speed as a function of frequency for various values of $\alpha$ is depicted in Fig.~\ref{fig:example}.
Since the consistency of the two-field model requires the hierarchy $M \ll \Lambda$, for concreteness we can imagine an example where $M=10^{-3}\Lambda$, so that the partial UV completion \eqref{eqn:UVscalar} remains a valid description up to the scale $\Lambda_*=10^3 \Lambda$. In that case if we were to draw an analogy with the frequencies observed at LIGO (\ie starting at about 24Hz), and considering the scale $\Lambda$ to be given by about $260$Hz as in eqn.~\eqref{eq:LambdaHorn}, then $k_{\rm LIGO}> 10^{-1}\Lambda\sim 10^2 M$, and we clearly see from Fig.~\ref{fig:example} that at sufficiently high frequencies, we expect the sound speed to be arbitrarily close to luminal, despite the low-frequency sound speed being potentially significantly subluminal.
It is worth noting that these scales should be taken with a grain of salt---they are merely provided to illustrate the point in this simple scalar field model and the precise way the sound speed returns to being luminal at high energy depends on the details of the (partial) UV completion. The purpose of this toy model is simply to illustrate that the measured GW speed at LIGO frequencies may be significantly different than the cosmologically relevant $c_T^2$---in practice, the precise numerical running and hierarchy between these two speeds will be determined by whatever physics UV completes Horndeski. \\

%%%%%%%%%%%%%%%%
\noindent{\bf Horndeski EFT:}
%%%%%%%%%%%%%%%%
%
We now turn to Horndeski as a dark energy EFT. As is well-known, the scalar field present in Horndeski can play the role of a dark energy fluid driving the late-time acceleration of the Universe. In doing so, the Universe is filled with a medium (the dark energy condensate) which in turn affects the GW speed.
% (without affecting those of the other massless particles such as photons).
For illustration purposes, consider the parts of the Horndeski dark-energy model which affects the sound speed \footnote{
Horndeski models of dark energy involve many other types of other operators which do not affect the sound speed. Including those would not affect our conclusions about $c_T$, namely that it is naturally of order unity near LIGO frequencies. However, these other operators may have other effects on the gravitational waveform, for example through a time-dependent Planck mass. If the spontaneous symmetry breaking is solely responsible for the time dependence of the Planck mass, then this will also asymptote to a constant in the UV as Lorentz invariance is restored. Interestingly, it was recently pointed out in \cite{Creminelli:2018xsv} that most Horndeski models of dark energy that do not predict a luminal GW sound speed would lead to a decay of GW into the dark energy field and are also severally constrained.
},
\ba
\label{eq:egHorndeski}
\mathcal{L}_{\rm H} &=&  \frac{\mpl^2}{2} R -\frac 12 \mathcal{G}^{ab} \p_a \phi \p_b \phi  \,, \\
 \mathcal{G}^{ab} &=& g^{ab}+c_2\frac{\mpl}{\Lambda^3}G^{ab}+c_3\frac{\mpl}{\Lambda^6}L^{a\mu b\nu}\nabla_\mu \nabla_\nu \phi\,,
\ea
$G_{ab}$ being the Einstein tensor and $L_{a\mu b\nu}$ the dual Riemann tensor, and we have defined the scale $\Lambda$ as $(H_0^2\mpl)^{1/3}$ as given in \eqref{eq:LambdaHorn}.
The solution $\langle \phi\rangle= \alpha \mpl H_0 t$ leads to an accelerated expansion with Hubble parameter $H=\beta H_0$, where the coefficients $\alpha$ and $\beta$ are determined in terms of $c_2$ and $c_3$ and are order one when $c_{2,3}$ are order one. There is a region in parameter space where the accelerated solutions are stable (no ghost nor gradient instabilities).
In order to exhibit the scales involved, it is useful to normalize metric fluctuations $g\mn = \gamma\mn+h\mn/\mpl$,
% on the background, the normalization of the scalar field perturbation $\delta \phi$ only depends only
% on the order one dimensionless coefficients $c_{2,3}$ and hence plays no role
so that the $c_{2,3} $ terms enter at the scale $\Lambda$,
\ba
\L_{\rm H} \supset (\p h)^2 +(\p \delta \phi)^2 + (\p \delta \phi)(\p \delta h) +\frac{1}{\Lambda^3} \p^2 h (\p \delta \phi)^2\! \! .
\quad\,\,
\ea
At first sight the $c_{2,3}$ terms in \eqref{eq:egHorndeski} would also seem to generate operators at a much lower scale, for instance  $\langle \dot \phi \rangle  \p^2 h \p \delta \phi/\Lambda^3\sim  \p^2 h \p \delta \phi / H_0 $, however all those operators are total derivatives.  \\

At low frequencies with respect to the cutoff $M$ of the Horndeski EFT, tensor modes have a subluminal speed,
\begin{equation}
 c_T^2 ( k ) =1-  \frac{2 c_2 \alpha^2\beta^2+6 c_3 \alpha^3 \beta^3}{2 +  c_2  \alpha^2\beta^2+6 c_3 \alpha^3 \beta^3}\, + \mathcal{O} \left(  \frac{k^2}{M^2} \right) \,,
\label{eq:cT0}
\end{equation}
where $M$  is at most the strong coupling scale of the EFT \footnote{A skeptic reader may worry about an EFT with such a low cutoff of the order of $10^{-13}$eV when GR is clearly valid and predictive over a much broader set of scales. Yet we should bear in mind that such a theory is typically introduced to tackle dark energy and would be valid from a scale of the order $10^{-33}$eV, that is at 20 orders of magnitude lower than that cutoff.},
but it may be lower, $M\lesssim \Lambda$ \footnote{
In theories that admit a Vainshtein mechanism \cite{Babichev:2013usa,Koyama:2013paa}, we may hope to be able to trust the theory at scales of order $\Lambda$ and to invoke a Vainshtein redressing to push the regime of validity of the theory to higher scales, however  the Vainshtein redressing  is negligible for the physical setup considered here.}.
%In what follows, we explore the possibility that Horndeski be a ``standard" effective field theory with a cutoff at or below $\Lambda$
\\

As was the case for the scalar field theory \eqref{eq:egPX}, the existence of a UV completion mandates the existence of other irrelevant operators in addition to the Horndeski ones. Precisely which operators would enter depends on the UV completion and within an EFT approach one should allow for all operators to be present. However, for concreteness, we present here a class of operators that would typically enter the Horndeski EFT at a scale $M\lesssim \Lambda$,
\ba
\L^{(n)}_{\rm higher-der}=\(\mpl^2 G_{\mu\nu}\) \frac{\Box^n}{M_n^{2n+4}}\p^\mu\phi\p^\nu\phi\,,
\label{eq:Ln}
\ea
with $n\ge 2$ and appropriate scales $M_n$, which we now study.
First, notice that such operators affect the background solutions by an amount proportional to
\ba
\frac{\mathcal{E}^{(n)}}{\mathcal{E}_{\rm H}}\sim \frac{H_0^{2(n-1)}\Lambda^6}{M_n^{2n+4}}\,,
\ea
where symbolically $\mathcal{E}^{(n)}$ is the contribution from $\L^{(n)}$ to the background equations of motion and $\mathcal{E}_{\rm H}$ that from the Horndeski Lagrangian \eqref{eq:egHorndeski}. Trusting the background provided by the Horndeski EFT \eqref{eq:egHorndeski} requires this ratio to be small.
So in principle the scale of the higher derivative operator $\L^{(n)}$ could be as small as say $M_n^{2n+4}\sim H_0^{2n-4}\Lambda^8 \lll \Lambda^{2n+4}$ and these operators would still not significantly affect the background.
Furthermore, on this background the higher derivative terms \eqref{eq:Ln} lead to operators that scale at worst as
$(\p^{n+1}h)^2 \p \delta \phi / \(H_0 M_n^{2n+4}\Lambda^{-3} \)$,
 (for $n\ge 2$),
so if those were at all representative of the types of operators we would expect from the UV completion, it would mean that the Horndeski EFT \eqref{eq:egHorndeski} can be trusted until the strong coupling scale $\Lambda_*$,
\ba
\Lambda_*={\rm min}_n \(M_n^{2n+4}H_0\Lambda^{-3}\)^{1/(2n+2)}\,.
\ea
It  will depend on the precise UV completion whether all the $M_n$ are the same order (maybe all set to $\Lambda$ or a lower scale $M$) or whether they scale so that $\Lambda_*> \Lambda$. For now we simply point out that we have a great deal of flexibility in the scales $M_n$ which do not alter the background evolution, yet {\bf do} affect the GW speed. For instance
\ba
\L=\L_{\rm H}+\sum_{n\ge 2}c_n \L^{(n)}_{\rm higher-der}
\ea
modifies the GW dispersion relation (symbolically) as,
\ba
&& \omega^2 \sim c_T^2 (0) k^2 +\mathcal{O}(H_0^2)\\
&+&\sum_{n\ge 2}\frac{c_n \Lambda^6}{3M_n^{2n+4}}\(-\omega^2+k^2\)^{n-1}(\omega^2+\mathcal{O}(k^2,H_0^2))\,,\nn
\ea
where at frequencies close to $M_n$ the $\(-\omega^2+k^2\)^{n-1}$ terms push the GW speed arbitrarily close to unity.
The rate at which the low energy sound speed asymptotes to luminal depends on the scales $M_n$, and is thus rather sensitive to details of the underlying UV completion. If one imagines a running in the form of a power law, $1/k^2$, then one requires that the cutoff of the theory is some orders of magnitude below the LIGO band if one is to accommodate $c_s \ll 1$ at low energies, but in principle the rate could be exponential or arbitrarily fast without affecting the low energy EFT.
 \\

%%%%
\noindent {\bf Conspiracy vs Lorentz-invariant UV completion:}
%%%%
The fact that the Horndeski cutoff is close to the LIGO band (and particularly the GW170817 event) was noticed in \cite{Creminelli:2017sry}, who pointed out that \emph{from a bottom-up approach} it would seem unlikely that order one effects entering at the cutoff would conspire to precisely cancel $c_T^2-1$ within an accuracy of one part in the $10^{15}$.
%From the point of view of the low-energy EFT this would indeed appear surprising.
However \emph{from a top-down approach}, it is very unlikely that the UV completion knows anything about the special structure of the sLB background. Quite the opposite, we expect that at sufficiently large energies modes should be insensitive to the sLB cosmological solution and we would naturally expect a return to luminality. Indeed the operators presented in \eqref{eq:Ln} (and \eqref{eqn:UVscalar}) have in no way been tuned so as to precisely cancel $c_T^2-1$. Rather the operators simply satisfy LI and at sufficiently high energy that symmetry is restored.
%What we mean by ``sufficiently'' high energy depends on the context, but for the GW170817 event, it is not inconceivable (and one may even argue natural) that the pure Horndeski theory (eg. \eqref{eq:egHorndeski}) has broken down by that scale and the speed of gravitational waves for those observed  frequencies has already returned to unity.
It is important to note that for the GW speed to be unity at LIGO frequencies, the EFT must breakdown at scales lower that $\Lambda$.   \\

%%%%
\noindent {\bf Modified Gravity:}
%%%%
One motivation for studying Horndeski is that these scalar-tensor theories can mimic  the behaviour of some modified gravity models \cite{Nicolis:2008in}: for instance the decoupling limit of DGP \cite{Chow:2009fm}, cascading gravity \cite{Agarwal:2009gy} and massive gravity \cite{deRham:2011by}.
%The initial study of Galileon scalar fields in \cite{Nicolis:2008in} was indeed motivated as a way to capture some of the relevant physics of infrared models of modified gravity.
%
Since some Horndeski EFTs arise from the decoupling limit of various theories of modified gravity, it is clear that Horndeski can be seen as an EFT with an infrared cutoff (of the order of the Hubble parameter today), as well as a UV cutoff and we could take the perspective that these models of modified gravity are in fact what (partially) ``completes" those Horndeski theories. Interestingly in all these models of modified gravity, while the dispersion relation is modified at very low frequencies (of the order of the effective graviton mass), the sound speed remains luminal independently of the background configuration. This suggests that Horndeski EFTs could very easily be implemented within some completion for which the GW speed  at LIGO frequencies is luminal to impeccable precision. All such EFTs may remain viable in the wake of GW170817.\\

%%%%
\noindent{\bf Gravitational Rainbows:}
%%%%
Throughout this work, we have raised the possibility that the frequencies observed at LIGO are at the edge of (or even beyond) the regime of validity of the  Horndeski EFT and shown how the speed of GWs could be close to unity at those scales even though the low-energy EFT may predict a subluminal propagation. By no means do we suggest that every time an observation is performed, one should simply shield the EFT from constraints by invoking a lower cutoff. However, within the context of Horndeski and current LIGO observations, the frequencies observed are dangerously close to the cutoff if the EFT is to describe dark energy and in a standard EFT approach new physics is \emph{required} to enter at or below that scale. \\

Turning towards future surveys, the upcoming LISA mission will have peak sensitivity near $10^{-3}$ Hz, at which scale $k/\Lambda\sim 10^{-5}$.
If LISA were to bound the speed of GWs \footnote{
Interestingly gas bounds to Black Holes in the LISA band may produce an X-ray signal \cite{Haiman:2017szj}.} with a similar precision as LIGO but at such low frequencies, it would be very hard for a Horndeski EFT to remain viable as a model of dark energy and still have an interesting regime of predictability.  Such observations would be complementary to those from future ground-based interferometers like the Einstein Telescope \cite{Sathyaprakash:2012jk} that may help distinguish between various dark energy models \cite{Belgacem:2017ihm,Belgacem:2018lbp}.  \\

Interestingly, in the case where $M$ is not much smaller than $\Lambda$, the running of $c_T^2 (k)$ induced by EFT corrections may be sufficiently large to rule out these models without the need for an optical counterpart.
The modification to the dispersion relation within the LIGO window would be dramatic, unless the transition between the low-energy and high-energy values of $c_T^2(k)$ happens extremely fast. If not, then for the example provided for Horndeski, it would require the higher derivative operators to enter at a scale at least 9 orders of magnitude below the observed scale so that we have completely transitioned between the low energy and high energy speed before LIGO starts taking data. \\

%%%%
\noindent{\bf Outlook for the EFT of Dark Energy:}
%%%%
In one of its simplest formulations \cite{Gleyzes:2013ooa}, the EFT of dark energy has only four free functions of time \footnote{In addition to those that determine the background cosmological history.}. One of those free functions ($m_4$) is directly related to the GW speed. While recent observations have been very successful at reducing the large parameter space, through this work we stress that those quantities are typically \emph{scale-dependent} (in addition to their time dependence) and the current constraints ($m_4(k_{\rm LIGO}) \approx 0$) may not necessarily imply $m_4(k\sim H_0\lesssim 10^{-20} k_{\rm LIGO})=0$.   \\

In particular, we have focused on a picture where new physics enters the low-energy EFT at a scale below $\Lambda = 260$Hz so as to restore perturbative unitarity. We should stress that even if the UV completion were to be {\it manifestly} Lorentz-violating, one would not expect the scale of Lorentz breaking at high energy to be linked to the scale of sLB at low energy and thus we would still expect a running of the speed of GWs.   \\

We emphasize that the aim of this work is not to revive Horndeski or any specific EFT as a particular model for dark energy. Rather the aim is to bring across the subtleties related with measurements such as the sound speed when dealing with EFTs, especially when the effective cutoff may be relatively low and comparable to the scale associated with the measurement.
In the coming age of precision cosmology, correctly interpreting what EFT corrections mean for these measurements will be more important than ever before and crucial for discriminating between different classes of models.

%\vspace{1cm}
\noindent {\bf Acknowledgments:}
We thank Paolo Creminelli, Lavinia Heisenberg, Atsushi Naruko, Andrew Tolley, Filippo Vernizzi and Toby Wiseman for useful discussions. CdR would like to thank the  Graduate Program on Physics for the Universe at Tohoku University for its hospitality during the latest stages of this work.
The work of CdR is supported by an STFC grant ST/P000762/1. CdR thanks the Royal Society for support at ICL through a Wolfson Research Merit Award. CdR is also supported in part by the European Union's Horizon 2020 Research Council grant 724659 MassiveCosmo ERC-2016-COG and in part by a Simons Foundation award ID 555326 under the Simons Foundation's Origins of the Universe initiative, `{\it Cosmology Beyond Einstein's Theory}'.
SM is funded by the Imperial College President's Fellowship.

\bibliography{references}

\end{document}